\documentclass[12pt,osajnl,letterpaper,showpacs,showkeys,groupedaddress,floatfix]{revtex4}
\usepackage[usenames]{color}
\usepackage[colorlinks,urlcolor=blue]{hyperref}
\usepackage{amssymb}
\usepackage{amsmath}
\usepackage{amsfonts}
\usepackage{bm}
\usepackage[dvips]{epsfig}
\usepackage{graphicx}
\providecommand{\U}[1]{\protect\rule{.1in}{.1in}}

\begin{document}
\title{Recursive $T$ matrix algorithm for resonant multiple scattering :\\Applications to localized plasmon excitations }
\author{Brian Stout$^{1}$ and J.C. Auger$^{2}$ and Alexis Devilez$^{1}$}
\affiliation{$^{1}$Case 161, Institut Fresnel, Facult\'{e} des Sciences et Techniques de
St. J\'{e}r\^{o}me }
\affiliation{13397 Marseille cedex 20, France }
\affiliation{$^{2}$Center for Laser Diagnostics, Dept. of Applied Physics }
\affiliation{Yale University, New Haven, CT 06520 USA}
\email{brian.stout@fresnel.fr,\ augerjc@gmail.com,\ alexis.devilez@fresnel.fr}
\date{July 2008}

\begin{abstract}
A matrix balanced version of the Recursive Centered $T$ Matrix Algorithm
(RCTMA) applicable to systems possessing \textit{resonant} inter-particle
couplings is presented. Possible domains of application include systems
containing interacting localized plasmon resonances, surface resonances, and
photonic jet phenomena. This method is of particular interest when considering
modifications to complex systems. The numerical accuracy of this technique is
demonstrated in a study of particles with strongly interacting localized
plasmon resonances.
\end{abstract}


\maketitle



\section{Introduction}

It has been well established that certain kinds of recursive $T$ matrix
algorithms (known as RCTMA)\cite{st02,au03} are numerically stable and can be
used to solve the Foldy-Lax multiple-scattering equations for particles
exhibiting \textquotedblleft modest\textquotedblright\ inter-particle
couplings. By \textquotedblleft modest couplings\textquotedblright,\ we refer
to situations in which the order of orbital number of the Vector Spherical
Wave Functions (VSWFs) necessary to describe the field scattered by each
particle in an aggregate of particles are not too much larger than that necessary
for describing isolated particles. The \textquotedblleft
modest coupling\textquotedblright\ criteria apply to a host of multiple
scattering situations, including systems of dielectric particles comparable in
size to the wavelength and for most packing fractions including dense packing.
The modest coupling criteria can also apply to metallic particles under
certain conditions.

Like any multiple scattering technique not employing matrix balancing, the RCTMA can
encounter numerical difficulties in certain extreme situations of strongly coupled resonant
phenomenon. In this work, we present a matrix balanced form of the Recursive Centered $T$
Matrix Algorithm (or RCTMA) that can readily be employed even in the presence
of strong (\textit{i.e.} resonant) inter-particle couplings. The rather
extreme situation of \textquotedblleft strong couplings\textquotedblright
\ studied here will generally require carefully micro-scaled engineered
systems where high $Q$-factor resonances can occur for particles illuminated
in isolation, and in which the particles are sufficiently closely spaced that
neighboring particles modify the resonance response properties. Examples of
strong inter-particle couplings can be found in particles exhibiting plasmon 
resonances, surface resonances, or even photonic jet phenomenon.

In section \ref{General}, the notation is introduced in a brief review of the
relevant multiple-scattering theory. Section \ref{Norm} describes an analytic matrix
balancing procedure used to `well-condition' the multiple scattering system of equations.
A matrix balanced RCTMA is derived in section \ref{modRCTMA}. Essential formulas
for applications are summarized in section \ref{Resume}. Their applications are then
demonstrated by applying matrix balanced RCTMA calculations to study systems of interacting 
localized plasmon excitations. Some known and novel aspects of interacting 
localized plasmon excitations are presented herein.

\section{\label{General}Multiple-scattering theory - VSWF approach}

\begin{figure}[ht]
\begin{center}
\includegraphics[width=0.5\textwidth]{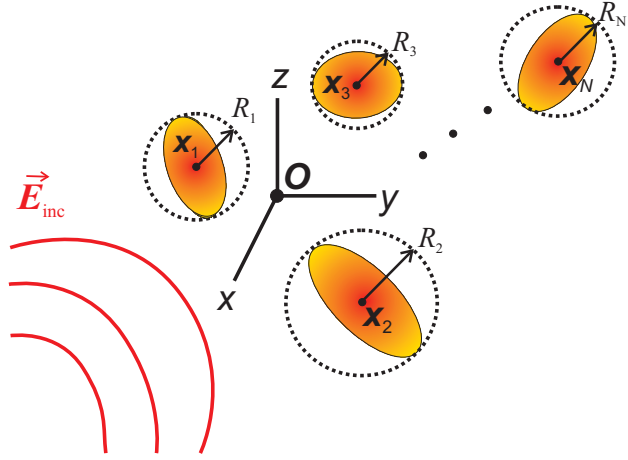}
\end{center}
\caption{\footnotesize Schematic of an field incident on a collection of scatterers centered
on $\mathbf{x}_{1},\mathbf{x}_{2},...,\mathbf{x}_{N}$. The radii of the
respective circumscribing spheres are denoted $R_{1},R_{2},...,R_{N}$.}
\label{fig:aggr}
\end{figure}

Let us consider an arbitrary incident electromagnetic
field incident on a collection of three-dimensional particles (as shown in
fig.\ref{fig:aggr}). The particles are considered as `individual' scatterers
if they can be placed in a circumscribing sphere lying entirely within
the homogeneous medium (actually this constraint can frequently be relaxed, 
cf. \cite{do06}).

The electromagnetic field incident on an $N$-particle system, $\mathbf{E}_{\mathrm{i}}$,
is developed in terms of the transverse \emph{regular} VSWFs developed about some point
$\mathbf{O}$ arbitrarily chosen as the system `origin':
\begin{align}
\mathbf{E}_{\mathrm{i}}\left(  \mathbf{r}\right)   &  =E_{0}\sum_{n=1}
^{\infty}\sum_{m=-n}^{n}\left\{  Rg \left[  \mathbf{M}_{nm}\left(
k\mathbf{r}\right)  \right]  a_{1,n,m}+Rg\left[ \mathbf{N}_{nm}\left(
k\mathbf{r}\right)  \right]  a_{2,n,m}\right\} \nonumber\\
&  = E_{0}\sum_{q=1}^{2}\sum_{p=1}^{\infty}Rg\left[ \mathbf{\Psi}
_{q,p}\left(  k\mathbf{r}\right)  \right]  a_{q,p}
 \equiv E_{0}\,Rg\left[  \mathbf{\Psi}^{t}\left(  k\mathbf{r}\right)
\right]  \,a \label{wavexp}
\end{align}
where $E_{0}$ is a real parameter determining the incident field amplitude. 
Eq.(\ref{wavexp}), introduces a condensed notation for the VSWFs, 
$\mathbf{M}_{nm}$ and $\mathbf{N}_{nm}$: 
$\mathbf{\Psi}_{1,p}\left(  k\mathbf{r}\right)  \equiv\mathbf{M}_{n,m}\left(
k\mathbf{r}\right)  $ and $\mathbf{\Psi}_{2,p}\left(  k\mathbf{r}\right)
\equiv\mathbf{N}_{n,m}\left(  k\mathbf{r}\right)$. The notation 
$Rg\left[\ \right]$ stands for \textquotedblleft the regular part
of\textquotedblright\ and distinguishes these regular VSWFs  from the
\textquotedblleft irregular\textquotedblright\ scattered VSWFs 
(cf. appendix \ref{VSH}). In the second line of eq.(\ref{wavexp}), the two 
subscripts $(n,m)$ are replaced by a single subscript $p$ defined such that 
$p\left(  n,m\right)  \equiv n(n+1)-m$ and has the inverse relations\cite{Ts85}:
\begin{equation}
n(p)=\mathrm{Int}\sqrt{p}\qquad\qquad m(p)=-p+n(n+1)\ .
\end{equation}
The last line of eq.(\ref{wavexp}), adopts the compact matrix notation allowing 
the suppression of the summation symbols.\cite{Ch94} The superscripted 
( $^{t}$ ) stands for the transpose of a column `matrix' of composed of VSWFs
into a row `matrix' of these functions.

For points external to all individual circumscribing spheres, the total field, 
$\mathbf{E}_{\mathrm{t}}\left(\mathbf{r}\right)$ can be written as the sum 
of the incident field, and a set of `individual' scattered fields, 
$\mathbf{E}_{\mathrm{s}}^{(j)}$, centered respectively on each of the particle centers:
\begin{align}
\mathbf{E}_{\mathrm{t}}\left(  \mathbf{r}\right)  &=\mathbf{E}_{\mathrm{i}}
\left(  \mathbf{r}\right)  +\sum_{j=1}^{N}\mathbf{E}_{\mathrm{s}}
^{(j)}\left(  \mathbf{r}_{j}\right) \nonumber \\
& = E_{0}\,Rg\left[  \mathbf{\Psi}^{t}\left(  k\mathbf{r}\right)
\right] \,a + E_{0} \sum_{j=1}^{N} \mathbf{\Psi}^{t}\left(  
k\mathbf{r}_{j}\right)  \,f_{N}^{(j)}
\end{align}
where each scattered field, $\mathbf{E}_{\mathrm{s}}^{(j)}$, is developed, 
with coefficients $f_{N}^{(j)}$, on a basis of outgoing VSWFs defined with 
respect to the associated particle center, denoted $\mathbf{x}_{j}$. The spherical coordinates 
relative to each scatterer are denoted $\mathbf{r}_{j}\equiv\mathbf{r-x}_{j}$.

The crucial idea of Foldy-Lax multiple-scattering theory is that there
exists an {\emph excitation field}, 
$\mathbf{E}_{\mathrm{exc}}^{\left(  j\right)}\left( \mathbf{r}\right)$, 
associated with each particle which is the superposition of the incident 
field and the field scattered by all the other
particles in the system (excluding the field scattered by the particle
itself).\cite{lax51} From this definition, the excitation field of the $j^{th}$
particle can be written
\begin{align}
\mathbf{E}_{\mathrm{exc}}^{\left(  j\right)  }\left(  \mathbf{r}_{j}\right)
&  \equiv E_{0}\,Rg\left[  \mathbf{\Psi}^{t}\left(  k\mathbf{r}_{j}\right)
\right] \,e_{N}^{(j)}\equiv\mathbf{E}_{\mathrm{i}}\left(  \mathbf{r}\right)
+\sum_{l=1,l\neq j}^{N}\mathbf{E}_{\mathrm{s}}^{(l)}\left(  
\mathbf{r}_{l}\right) \nonumber\\
&  =E_{0}\, Rg\left[  \mathbf{\Psi}^{t}\left(  k\mathbf{r}_{j}\right)  \right]
\left[  J^{(j,0)}a+\sum_{l=1,l\neq j}^{N}H^{(j,l)}\,f_{N}^{(l)}\right]
\label{excdef}
\end{align}
where $e_{N}^{(j)}$ are the coefficients of the excitation field in a regular 
VSWF basis centered on the $j^{th}$ particle. In the second line of eq.(\ref{excdef}), we
have used the translation-addition theorem\cite{Ts85,Ch94,st02}) and introduced the notation
where $J^{(j,0)}\equiv$ $J\left(  k\mathbf{x}_{j}\right)$ is a regular
translation matrix and $H^{(j,l)}\equiv H\left[  k\left(  
\mathbf{x}_{j}-\mathbf{x}_{l}\right) \right]$ is an irregular translation matrix.
Analytical expressions for the matrix elements of 
$J\left( k\mathbf{x}_{j}\right)$ and $H\left(  k\mathbf{x}_{j}\right)$ are 
given in refs.\cite{Ts85,st02}.

The other key idea of multiple scattering theory is that the field scattered
by the object, $f_{N}^{(j)}$, is obtained from the excitation field
$e_{N}^{(j)}$ via the 1-body $T$ matrix, $T_{1}^{\left(  j\right)}$, derived
when one considers the particle to be immersed in an infinite homogeneous
medium. This relation is then expressed as
\begin{equation}
f_{N}^{(j)}=T_{1}^{\left(  j\right)}e_{N}^{(j)} \label{T1def}
\end{equation}
[The index $1$ on the $T_{1}^{\left(  j\right)  }$ indicates that this $T$
matrix concerns an isolated particle, henceforth referred to as a `1-body' $T$
matrix.] Employing eq.(\ref{T1def}) in eq.(\ref{excdef}), one obtains a
Foldy-Lax set of equations for the excitation field
coefficients\cite{st02,Ch94}:
\begin{equation}
e_{N}^{(j)}=J^{(j,0)}a+\sum_{l=1,l\neq j}^{N}H^{(j,l)}\,T_{1}^{\left(
l\right)  }\,e_{N}^{(l)} \qquad j=1,...,N \label{excFoldy}
\end{equation}

For numerical applications where one is obliged to solve the equations on a
truncated VSWF basis, it is advantageous to work with a set of formally
equivalent equations involving the scattering coefficients $f_{N}^{(j)}$. This
set of equations is derived by multiplying each of eqs.(\ref{excFoldy})
from the left by $T_{1}^{(j)}$ and using eq.(\ref{T1def}) to obtain
\begin{equation}
f_{N}^{(j)}=T_{1}^{(j)}J^{(j,0)}a+T_{1}^{(j)}\sum_{l=1,l\neq j}^{N}
H^{(j,l)}\,f_{N}^{(l)} \qquad j=1,...,N  \label{fFoldy}
\end{equation}

In the RCTMA, one calculates the \emph{centered} multiple scattering 
transition matrices, $T_{N}^{(j,k)}$, which directly yield the
scattered field coefficients in terms of the field incident on the system
through the expression
\begin{equation}
f_{N}^{(j)}=\sum_{k=1}^{N}T_{N}^{(j,k)}\,a^{(k)}\qquad\text{with}\qquad
a^{(k)}\equiv J^{(k,0)}\,a \label{Tjl}
\end{equation}
In this equation, we have introduced the column matrix $a^{(j)}$ which contains the
coefficients of the field incident on the entire system developed on a VSWF
basis centered on the $j^{th}$ particle. 

\section{\label{Norm}Basis set truncation and matrix balancing}

Although the multiple scattering formulas of the previous section are 
expressed as matrix equations on VSWF basis sets of
infinite dimension, the finite size of the scatterers naturally
restricts the dimension of the dominate VSWF contributions.
In order to discuss this phenomenon analytically, we consider the case of
spherical scatterers. For non-spherical scatterers, the matrix balancing
procedure described below should be applied to the circumscribing spheres
of the particles.

The Mie solution for a sphere of radius $R_{j}$ immersed in a homogeneous
host medium, can be cast in the form of a 
1-body $T$ matrix that is diagonal in a VSWF basis centered on the particle:
\begin{equation}
\left[  T_{1}^{\left(  j\right)  }\right] _{q,p;q^{\prime},p^{\prime}}
=\delta_{q,q^{\prime}}\delta_{p,p^{\prime}}T_{1}\left( j,n(p),q\right) \label{diag} 
\end{equation}
where the $T_{1}^{\left(  j\right) }\left(  n(p),q\right)$ correspond to the Mie
coefficients and depend on $q$ and $n$ (cf. eq.(\ref{wavexp})).

\begin{figure}[ht]
\begin{center}
\includegraphics[width=0.5\textwidth]{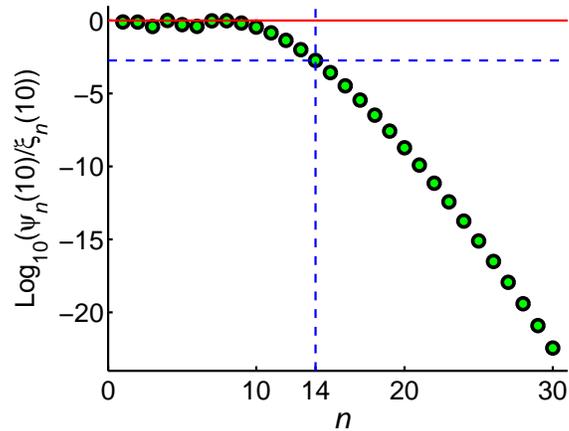}
\end{center}
\caption{\footnotesize Plot of the spherical Bessel to Hankel function ratio, 
$\psi_{n}\left(  kR\right) /\xi_{n}\left(  kR\right)$ occurring in the
Mie coefficients when $kR=10$.}
\label{fig:norm}
\end{figure}

With the objective of matrix balancing, it is helpful to express the 
Mie coefficients of the scatterers in terms of the Ricatti Bessel and Hankel 
functions, respectively 
$\psi_{n}\left(  z\right)  \equiv zj_{n}\left(  z\right)$ and  
$\xi_{n}\left(z\right)  \equiv zh_{n}\left(  z\right)$,
and their logarithmic derivatives
\begin{equation}
\Phi_{n}\left(  z\right)  \equiv\frac{\psi_{n}^{\prime}\left(  z\right)}
{\psi_{n}\left(  z\right)  }\qquad\Psi_{n}\left(  z\right) \equiv\frac
{\xi_{n}^{\prime}\left(  z\right)  }{\xi_{n}\left(  z\right) }
\end{equation}
The $T$ matrix elements of eq.(\ref{diag}) for a sphere of dielectric contrast 
$\rho_{j}\equiv k_{j}/k$ can then be cast in the convenient form\cite{st03}:
\begin{align}
T\left(  j,n,1\right)   &  =\frac{\psi_{n}\left(  kR_{j}\right)  }{\xi
_{n}\left(  kR_{j}\right)  }\,\frac{\frac{\mu_{j}}{\mu}\Phi_{n}\left(
kR_{j}\right)  -\rho_{j}\Phi_{n}\left(  \rho_{j}kR_{j}\right)  }
{\rho_{j}\Phi_{n}\left(  \rho_{j}kR_{j}\right)  -\frac{\mu_{j}}{\mu}\Psi_{n}\left(
kR_{j}\right)  }\equiv\frac{\psi_{n}\left(  kR_{j}\right)  }{\xi_{n}\left(
kR_{j}\right)  }\,\overline{T}\left(  j,n,1\right) \nonumber\\
T\left(  j,n,2\right)   &  =\frac{\psi_{n}\left(  kR_{j}\right) }
{\xi_{n}\left(  kR_{j}\right)  }\,\frac{\frac{\mu_{_{j}}}{\mu}\Phi_{n}
\left(\rho_{_{j}}kR_{j}\right)  -\rho_{j}\Phi_{n}\left(  kR_{j}\right)}
{\rho_{_{j}}\Psi_{n}\left(  kR_{j}\right)  -\frac{\mu_{j}}{\mu}\Phi_{n}\left(
\rho_{j}kR_{j}\right)  }\equiv\frac{\psi_{n}\left(  kR_{j}\right) }
{\xi_{n}\left(  kR_{j}\right)  }\,\overline{T}\left(  j,n,2\right)  \label{MieT}
\end{align}
where $k$ is the wavenumber in the external medium. The {\emph normalized}
$T$ matrix coefficients, $\overline{T}\left(  j,n,q\right)$, of eq.(MieT) contain a rich 
resonant structure. The ratios
$\psi_{n}\left(  kR_{j}\right) /\xi_{n}\left(  kR_{j}\right)$ on the other hand 
have an exponentially decreasing behavior for large, $n\gg kR_{j}$ as is demonstrated in 
fig.\ref{fig:norm} for $kR=10$. One can remark from figure \ref{fig:norm} that 
$\left\vert \psi_{n}\left(  kR\right)/\xi_{n}\left(kR\right)\right\vert$
become quite small beyond $n_{\mathrm{\max}}=kR+3$ and its value at
$n=14$ is $\sim2\,10^{-4}$. Although these factors permit an appropriately
truncated VSWF basis set to contain essentially all the physical information necessary
for accurate calculations, they also tend to produce ill-conditioned linear systems 
when one is obliged to enlarge the VSWF space far beyond $\approx kR+3$ in 
order to account for strong coupling phenomenon.

A solution to the above problem is to `balance' the matrix manipulations in
section \ref{modRCTMA} below by defining \textquotedblleft
normalized\textquotedblright\ scattering and incident coefficients:
\begin{align}
\left[  \overline{f}^{\left(  j\right) }\right]_{q,p}  &  
\equiv \xi_{n\left(  p\right)}\left(  kR_{j}\right)  \left[  
f^{\left( j\right) }\right]_{q,p} \nonumber\\
\left[  \overline{a}^{\left(  j\right)  }\right]_{q,p}  &  \equiv
\psi_{n\left(  p\right)  }\left(  kR_{j}\right)  \left[  a^{\left(  j\right)
}\right]  _{q,p} \label{normcoeffs}
\end{align}
For notational purposes, it is convenient to define diagonal matrices
$\left[ \psi^{\left( j\right)}\right]$ and $\left[\xi^{\left( j\right)}\right]$ 
with Ricatti-Bessel functions along their diagonals, namely $\left[  \psi^{\left(  j\right)
}\right]  _{q^{\prime},p^{\prime};q,p}$ $\equiv\delta_{q,q^{\prime}}
\delta_{p,p^{\prime}}\psi_{n(p)}\left(  kR_{j}\right)  $ and $\left[
\xi^{\left(  j\right)  }\right]_{q^{\prime},p^{\prime};q,p}$ $\equiv
\delta_{q,q^{\prime}}\delta_{p,p^{\prime}}\xi_{n(p)}\left(  kR_{j}\right)$.
This notation allows normalized or `balanced' versions 
of the one-body and many-body $T$-matrices to be defined respectively as
\begin{equation}
\overline{T}_{1}^{(j)}\equiv\left[  \xi^{\left(  j\right)  }\right]
T_{1}^{(j)}\left[  \psi^{\left(  j\right)  }\right]^{-1} 
\qquad {\rm and} \qquad \overline{T}_{N}^{(j,k)}\equiv \left[  \xi^{\left(  j\right)
}\right]  \overline{T}_{N}^{(j,k)}\,\left[  \psi^{\left(  k\right)  }\right]
^{-1} \label{Tnorm}
\end{equation}
In terms of these normalized quantities, eq.(\ref{Tjl}) then reads
\begin{equation}
\overline{f}_{N}^{(j)}=\sum_{k=1}^{N}\overline{T}_{N}^{(j,k)}\,\overline
{a}^{\left(  k\right)  }  \qquad\qquad j=1,...,N \label{Npartsol}
\end{equation}
In the next section, these `normalized'\, $\overline{T}_{1}^{(j)}$ 
and $\overline{T}_{N}^{(j,k)}$ are used to derive a matrix balanced version of the
recursive $T$ matrix algorithm.

\section{\label{modRCTMA}Derivation of a matrix balanced recursive algorithm}

In this section, we derive a matrix balanced version of the Recursive Centered 
T Matrix Algorithm (RCTMA) using purely algebraic manipulations. The recursive 
algorithm can be invoked once we have a
solution for the $\overline{T}_{N-1}^{(j,k)}$ matrices of a $N\geq1$ particle
system. If we wish to solely use the recursive algorithm to solve a system, we 
initiate the recursive process with a single particle solution described by
$\overline{T}_{1}^{(1,1)}\equiv\overline{T}_{1}^{\left( 1\right)}$.

One then considers an arbitrarily positioned particle being added to the
system. The excitation field on a particle $N$ added to the system can be
expressed as the superposition of three fields. The first contribution is simply the
field incident on the system, the second contribution results from the scattering of
the incident field by the $N-1$ cluster of particles onto the particle $N$,
and finally the third contribution comes from field scattered by the particle
$N$ onto the $N-1$ cluster and which returns to the $N^{th}$ particle as an
excitation field. Invoking the translation-addition theorem and eq.(\ref{Tjl}),
these three contributions can be expressed in matrix form as\cite{st02}
\begin{equation}
e_{N}^{(N)}=a^{\left(  N\right)  }+\sum_{j,k=1}^{N-1}H^{\left(  N,j\right)}
\,T_{N-1}^{(j,k)}\,a^{\left(  k\right)  }+\sum_{j,k=1}^{N-1}H^{\left(
N,j\right)  }T_{N-1}^{(j,k)}\,H^{\left(  k,N\right) }\,f_{N}^{(N)}
\label{nonorm}
\end{equation}

Defining now the normalized irregular translation matrices and excitation coefficients 
respectively as
\begin{equation}
\overline{H}^{\left(  j,k\right)  }\equiv\left[  \psi^{\left( j\right)
}\right]  H^{\left(  j,k\right)  }\left[  \xi^{\left(  k\right) }\right]
^{-1}  \qquad {\rm and} \qquad \overline{e}_{N}^{(j)}\equiv \left[ \psi^{\left(  j\right)
}\right] e_{N}^{\left(  j\right)  }
\label{enorm}
\end{equation}
the normalized form of eq.(\ref{nonorm}) can be written
\begin{equation}
\overline{e}_{N}^{(N)}=\overline{a}^{\left(  N\right)  }+
\sum_{j,k=1}^{N-1}\overline{H}^{\left(  N,j\right)}\overline{T}_{N-1}^{(j,k)}
\,\overline{a}^{\left(  k\right)  }+\sum_{j,k=1}^{N-1}\overline{H}^{\left(
N,j\right)  }\overline{T}_{N-1}^{(j,k)}\,\overline{H}^{\left(  k,N\right)}
\overline{f}_{N}^{(N)} \label{excN}
\end{equation}
where we also used the definitions in eqs.(\ref{normcoeffs}) and
(\ref{Tnorm}).

Recalling that the excitation field is linked to the scattered field by the
1-body $T$-matrix via eq.(\ref{T1def}), and invoking the definitions of
eq.(\ref{normcoeffs}) and (\ref{enorm}) we can write
\begin{equation}
\overline{e}_{N}^{(j)}=\left[  \overline{T}_{1}^{(j)}\right]^{-1}
\overline{f}_{N}^{(j)} \label{normexc}
\end{equation}
Employing this relation for particle $N$ on the LHS of eq.(\ref{excN}) and
rearranging we obtain
\begin{align}
\left\{  \left[  \overline{T}_{1}^{(N)}\right]^{-1} -\sum_{j,k=1}
^{N-1}\overline{H}^{\left(  N,j\right)  }\overline{T}_{N-1}^{(j,k)}
\,\overline{H}^{\left(  k,N\right)  }\right\}  \,\overline{f}_{N}
^{(N)} & \nonumber\\
& \hspace{-200pt} =\overline{a}^{\left(  N\right)  }+\sum_{j,k=1}^{N-1}\overline
{H}^{\left(  N,j\right)  }\,\overline{T}_{N-1}^{(j,k)}\,
\overline{a}^{\left(k\right)} \label{fN}\ .
\end{align}
We now take the normalized $\overline{T}_{N}^{(N,N)}$ matrix to be expressed
as
\begin{equation}
\overline{T}_{N}^{(N,N)}=\left\{  \left[  \overline{T}_{1}^{(N)}\right]
^{-1}-\sum_{j,k=1}^{N-1}\overline{H}^{\left(  N,j\right)  }
\overline{T}_{N-1}^{(j,k)}\,\overline{H}^{\left(  k,N\right)}\right\}^{-1}
\label{TNNnew}
\end{equation}
With this assignment, we multiply both sides of eq.(\ref{fN}) by $\overline
{T}_{N}^{(N,N)}$ and obtain an expression consistent with equation
(\ref{Npartsol}):
\begin{align}
\overline{f}_{N}^{(N)}  &  =\overline{T}_{N}^{(N,N)}\,\overline{a}^{\left(
N\right)  }+\overline{T}_{N}^{(N,N)}\sum_{j,l=1}^{N-1}\overline{H}^{\left(
N,j\right)  }\,\overline{T}_{N-1}^{(j,k)}\,\overline{a}^{\left(  k\right)}
\nonumber\\
&  =\overline{T}_{N}^{(N,N)}\,\overline{a}^{\left(  N\right) }+
\sum_{l=1}^{N-1}\overline{T}_{N}^{(N,k)}\,\overline{a}^{\left(  k\right)}
=\sum_{l=1}^{N}\overline{T}_{N}^{(N,k)}\,\overline{a}^{\left(  k\right)  }
\label{fNres}
\end{align}
where we have assigned the matrix $\overline{T}_{N}^{(N,k)}$, $k\neq N$ as
\begin{equation}
\overline{T}_{N}^{(N,k)}=\overline{T}_{N}^{(N,N)}\sum_{j=1}^{N-1}
\overline{H}^{\left(  N,j\right)  }\,\overline{T}_{N-1}^{(j,k)} \label{TNk}
\end{equation}

One completes the description of the scattering by the system by remarking
that the field scattered by the other particles in the system are the
superposition of the field that would be scattered by the $N-1$ cluster in the
absence of the $N^{th}$ particle, plus the field scattered from the $N-1$
particle originating as a scattered field emanating from the $N^{th}$
particle. Using again the translation-addition theorem, the field coefficients
of $\overline{f}_{N}^{(j)}$ can in turn be expressed in a form consistent with
equation (\ref{Npartsol}) as
\begin{align}
\overline{f}_{N}^{(j)}  &  =\sum_{l=k}^{N-1}\overline{T}_{N-1}^{(j,k)}
\,\overline{a}^{\left(  k\right)  }+\sum_{k=1}^{N-1}\overline{T}_{N-1}
^{(j,k)}\,\overline{H}^{(k,N)}\,\overline{f}_{N}^{(N)}\nonumber\\
&  =\sum_{l=1}^{N-1}\overline{T}_{N-1}^{(j,k)}\,\overline{a}^{\left(
k\right)  }+\sum_{k=1}^{N-1}\overline{T}_{N-1}^{(j,k)}\,\overline{H}
^{(k,N)}\,\overline{T}_{N}^{(N,N)}\,\overline{a}^{\left(  N\right)  }
\nonumber\\ & \qquad +\sum_{l=1}^{N-1}\sum_{k=1}^{N-1}\overline{T}_{N-1}^{(j,l)}\,\overline
{H}^{(l,N)}\,\overline{T}_{N}^{(N,k)}\,\overline{a}^{\left(  k\right)
}\nonumber\\
&  =\overline{T}_{N}^{(j,N)}\,\overline{a}^{\left(  N\right)  }+\sum
_{k=1}^{N-1}\overline{T}_{N}^{(j,k)}\,\overline{a}^{\left(  k\right)  }
=\sum_{k=1}^{N}\overline{T}_{N}^{(j,k)}\,\overline{a}^{\left(  k\right) }
\end{align}
where we invoked eq.(\ref{fNres}). In the second and third lines 
we have defined the $\overline{T}_{N}^{(j,N)}$ and 
$\overline{T}_{N}^{(j,l)}$ matrices such that
\begin{subequations}
\label{Treadjust}
\begin{align}
\overline{T}_{N}^{(j,N)}  &  =\sum_{k=1}^{N-1}\overline{T}_{N-1}^{(j,k)}
\,\overline{H}^{(k,N)}\,\overline{T}_{N}^{(N,N)}\label{TjN}\\
\overline{T}_{N}^{(j,k)}  &  =\overline{T}_{N-1}^{(j,k)}+\sum_{l=1}^{N-1}
\overline{T}_{N-1}^{(j,l)}\,\overline{H}^{(l,N)}\,
\overline{T}_{N}^{(N,k)}
\end{align}
\end{subequations}

At this point, all the $\overline{T}_{N}^{(j,k)}$ matrices have been obtained
and the matrix manipulations in eqs.(\ref{TNNnew}), (\ref{TNk}) and
(\ref{Treadjust}) can then be repeated to add as many particles to the system
as desired.

\subsection{Relationship with system matrix inversions}

Although the recursive algorithm is quite efficient for systems with relatively
small numbers of particles, for systems with many particles, one may prefer to try 
and solve an entire $N$-particle system directly. A balanced linear system for the 
entire system corresponding to our recursive algorithm can be obtained by applying the
relation of eq.(\ref{T1def}) to the left hand side of eq.(\ref{fFoldy}), 
then multiplying both sides of the resulting equations by the 
$\left[\psi^{\left( j\right)}\right]$ matrix and finally rearranging to obtain a
system of balanced linear equations for the unknown scattering coefficients:
\begin{equation}
\left[ \overline{T}_{1}^{(j)}\right]^{-1}\overline{f}_{N}^{(j)}
-\sum_{k=1,k\neq j}^{N}\overline{H}^{\left(  j,k\right)  }\,\overline{f}
_{N}^{(k)}=\overline{a}^{\left(  j\right) }\qquad\qquad\qquad j=1,...,N
\label{linsyst}
\end{equation}
where we used eqs.(\ref{normcoeffs}) and (\ref{Tnorm}). The system of linear
equations in eq.(\ref{linsyst}) can in principle be directly solved by
inverting the balanced system matrix:
\begin{equation}
\left[
\begin{array}
[c]{c}
\overline{f}_{N}^{(1)}\\
\overline{f}_{N}^{(2)}\\
\vdots\\
\overline{f}_{N}^{(N)}
\end{array}
\right] =\left[
\begin{array}
[c]{cccc}
\left[ \overline{T}_{1}^{(1)}\right]^{-1} & -\overline{H}^{(1,2)} & \cdots
& -\overline{H}^{(1,N)}\\
-\overline{H}^{(2,1)} & \left[  \overline{T}_{1}^{(2)}\right]^{-1} & \cdots
& -\overline{H}^{(2,N)}\\
\vdots & \vdots & \ddots & \vdots\\
-\overline{H}^{(N,1)} & -\overline{H}^{(N,2)} & \cdots & \left[  \overline
{T}_{1}^{(N)}\right]  ^{-1}
\end{array}
\right]  ^{-1}\left[
\begin{array}
[c]{c}
\overline{a}_{{}}^{\left(  1\right)  }\\
\overline{a}_{{}}^{\left(  2\right)  }\\
\vdots\\
\overline{a}_{{}}^{\left(  N\right)  }
\end{array}
\right]  \label{matinv}
\end{equation}
Once we have inverted this system, one can associate each block with a
corresponding $\overline{T}_{N}^{(j,k)}$ matrix as shown below as
\begin{equation}
\left[
\begin{array}
[c]{c}
\overline{f}_{N}^{(1)}\\
\overline{f}_{N}^{(2)}\\
\vdots\\
\overline{f}_{N}^{(N)}
\end{array}
\right]  =\left[
\begin{array}
[c]{cccc}
\overline{T}_{N}^{(1,1)} & \overline{T}_{N}^{(1,2)} & \cdots & \overline
{T}_{N}^{(1,N)}\\
\overline{T}_{N}^{(2,1)} & \overline{T}_{N}^{(2,2)} & \cdots & \overline
{T}_{N}^{(2,N)}\\
\vdots & \vdots & \ddots & \vdots\\
\overline{T}_{N}^{(N,1)} & \overline{T}_{N}^{(N,2)} & \cdots & \overline
{T}_{N}^{(N,N)}
\end{array}
\right] \left[
\begin{array}
[c]{c}
\overline{a}_{{}}^{\left(  1\right)  }\\
\overline{a}_{{}}^{\left(  2\right)  }\\
\vdots\\
\overline{a}_{{}}^{\left(  N\right)  }
\end{array}
\right]  \label{system}
\end{equation}
which is the same form as the desired solutions given in eq.(\ref{Npartsol}).

\section{\label{Resume}Summary and applications to localized plasmon excitations}

In this section, we will apply the RCTMA to solve systems exhibiting strong interactions
between localized plasma resonances. We begin this section by summarizing the balanced recursive algorithm. 
We then recall some useful formulas for extracting physical quantities from the $T$ matrix.
Finally, we carry out some illustrative calculations for strongly interacting systems.

\subsection{Summary of the balanced RCTMA algorithm}

In order to implement the RCTMA, one must first solve the $1$-body $T$-matrices, 
$T_{1}^{\left(  1\right)  }$,$T_{1}^{\left(  2\right)  }$,$...$,
$T_{1}^{\left(  N_{\mathrm{tot}}\right) }$,
for all the particles in the system. Normalized versions of the $1$-body $T$ matrices 
and the irregular translation matrices\cite{Ts85,st02}, $H^{(j,k)}$, are
then calculated via
\begin{equation}
\overline{T}_{1}^{(j)}\equiv\left[  \xi^{\left(  j\right)  }\right]
T_{1}^{(j)}\left[  \psi^{\left(  j\right)  }\right]  ^{-1} \qquad 
\overline{H}^{\left(  j,k\right)  }\equiv\left[  \psi^{\left(  j\right)}\right]  
H^{\left(  j,k\right)  }\left[  \xi^{\left(  k\right)  }\right]^{-1}
\end{equation}
where the diagonal matrices $\left[  \psi^{\left(  j\right)  }\right]
_{q,q^{\prime},p,p^{\prime}}=\delta_{q,q^{\prime}}\delta_{p,p^{\prime}}
\psi_{n\left(  p\right)  }\left( kR_{j}\right)$ and $\left[  \xi^{\left(
j\right)  }\right]  _{q,q^{\prime},p,p^{\prime}}=\delta_{q,q^{\prime}}
\delta_{p,p^{\prime}}\xi_{n\left(  p\right)  }\left(  kR_{j}\right)$
respectively have Ricatti-Bessel and Ricatti-Hankel functions on the diagonal.
($R_{j}$ being the radius of the circumscribing sphere of the $jth$ scatterer).

The balanced recursive algorithm is that the solution for the $T_{N}^{(N,N)}$ matrix 
is obtained from the $T$ matrices of the $N-1$ system, $T_{N-1}^{(j,k)}$, via the 
matrix inversion in eq.(\ref{TNNnew}). All the other matrices 
$\overline{T}_{N}^{(j,k)}$ with $j\neq N$ or 
$k\neq N$ are then obtained via matrix multiplications and additions via 
equations (\ref{TNk}) and (\ref{Treadjust}). This process is then repeated as many
times as desired.

\subsection{Physical quantities}

When the incident field is a plane wave, 
it is convenient to express physical quantities in terms of cross sections.
Appealing to the far-field approximation of the field, the extinction
and scattering cross sections of clusters of $N$ objects can be respectively 
expressed\cite{st01,Ma94}
\begin{equation}
\sigma_{\mathrm{ext}}=-\frac{1}{k^{2}}\operatorname{Re}\left[  \sum_{k=1}^{N}
 a^{\left(  j\right)  ,\dagger}f_{N}^{\left(  j\right)  }\right] \qquad {\rm and}
\qquad  \sigma_{\mathrm{scat}}=\frac{1}{k^{2}}\sum_{j,k=1}^{N}f_{N}^{\left(  j\right)
,\dagger}J^{\left( j,k\right) }f_{N}^{\left(  k\right) }
\label{extscat}
\end{equation}

It is also possible to produce analytical expression for local
field quantities like individual absorption cross sections. 
For lossy scatterers in a lossless host medium, one can obtain  
individual particle absorption cross sections by 
integrating the Poynting vector on a circumscribing sphere surrounding the
particle to obtain the formula as
\begin{equation}
\sigma_{\mathrm{a}}^{(j)}=-\frac{1}{k^{2}}\mathrm{Re}
\left\{ f_{N}^{(j),\dagger}\, e_{N}^{(j)}\right\}  
-\frac{1}{k^{2}}\left\vert f_{N}^{(j)}\right\vert^{2}\label{abs}
\end{equation}

In an analogous fashion, optical forces on the particles can be calculated by
integrating the Maxwell tensor on a circumscribing sphere surrounding the
particle\cite{moine05,mis02}. It is frequently convenient to characterize 
the optical force by vector `cross sections',
$\overrightarrow{\sigma}_{\mathrm{opt}}$, defined such that the time averaged
optical force on particles immersed in a liquid dielectric of refraction index
$n_{\mathrm{med}}$ can be expressed as
\begin{equation}
\mathbf{F}_{\mathrm{opt}}=\left\Vert \mathbf{S}_{\mathrm{inc}}\right\Vert
\frac{n_{\mathrm{med}}}{c}\overrightarrow{\sigma}_{\mathrm{opt}}
\end{equation}
where $\left\Vert \mathbf{S}_{\mathrm{inc}}\right\Vert =\left\Vert 
\frac{1}{2}\operatorname{Re}\{\mathbf{E}_{\mathrm{inc}}^{\ast}\times\mathbf{H}
_{\mathrm{inc}}\}\right\Vert $ is the incident irradiance. The binding force
and its associated cross section, $\sigma_{\rm b}$, between two particles 
separated by a relative position vector 
$\mathbf{r}_{\mathrm{pos}}\equiv\mathbf{r}_{2}-\mathbf{r}_{1}$, can be 
defined as
\begin{equation}
F_{\mathrm{b}}\equiv\frac{1}{2}\left(  \mathbf{F}_{2}-\mathbf{F}_{1}\right)
\cdot\widehat{\mathbf{r}}_{\mathrm{pos}}\equiv\left\Vert 
\mathbf{S}_{\mathrm{inc}}\right\Vert \frac{n_{\mathrm{med}}}{c}\sigma_{\mathrm{b}}
\label{binding}
\end{equation}

\subsection{Interacting localized plasmon excitations}

For those conductors, such as the noble metals, that support surface 
plasmon resonances, one can usually observe localized plasmon resonances 
in sufficiently small particles. These resonances are typically dominated by 
absorption if the particles are sufficiently small with respect to the incident 
wavelength and by scattering for larger particles. We chose to study silver spheres 
$50\ \mathrm{nm}$ in diameter immersed in air (for which both scattering and absorption 
are non-negligible).

\begin{figure}[ht]
\centering\includegraphics[width=0.5\textwidth]{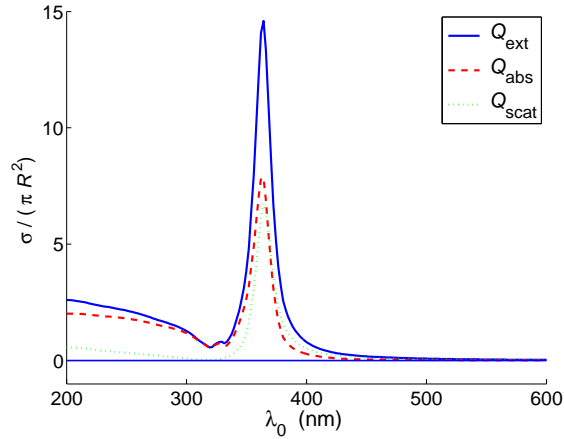}
\caption{\footnotesize Total cross section `efficiencies', $Q \equiv \sigma/(\pi R^{2})$ for an isolated 50 $\mathrm{nm}$
diameter sphere.}
\label{fig_cross_sphere}
\end{figure}

The study is carried out for wavelengths ranging from the near 
ultra-violet through the visible (300 to 850 $\mathrm{nm)}$. We ignore
the relatively modest finite size corrections to damping\cite{bohr83} and simply 
adopt the bulk dielectric constant of silver from ref.\cite{AIP} and extrapolate 
between the experimental values. The extinction, scattering and absorption 
cross sections for these particles are readily obtained from Mie theory 
and are displayed in figure \ref{fig_cross_sphere} as a function of frequency.
These spheres are quite small with respect to visible wavelengths, 
(size parameters in the 300$\leftrightarrow$800 nm wavelength range go through 
$kR=0.52\leftrightarrow0.20$) and the isolated particle cross
sections are obtained to high precision with $n_{\max}=4$. One can also
see from figure \ref{fig_cross_sphere} that the strength of the plasmon
resonance for these particles is about half due to absorption and
about half due to scattering.

\begin{figure}[ht]
\centering\includegraphics[width=0.5\textwidth]{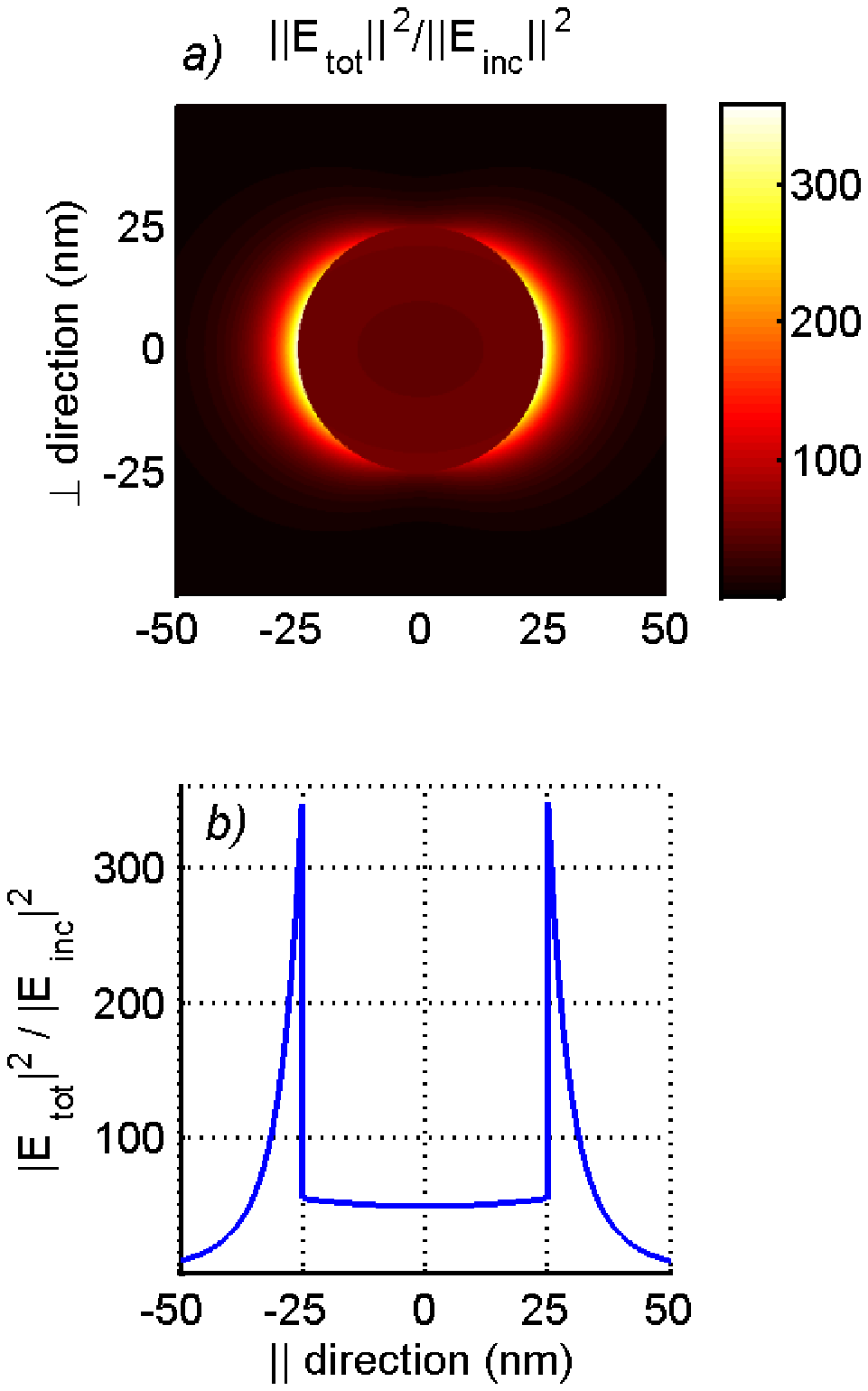}
\caption{{\footnotesize Electric field intensity 
$||\mathbf{E}_{\mathrm{t}}||^{2}/||\mathbf{E}_{\mathrm{inc}}||^{2}$ in an isolated 
$50\ \mathrm{nm}$ diameter sphere 
($\lambda_{0}=365\ \mathrm{nm}$, $N_{\mathrm{Ag}}=0.077+1.6i$). In a) is presented a 2D (hot) 
plot of the electric field intensity in a plane perpendicular to the wavevector and
containing the origin of the sphere (the horizontal axis lies along the polarization direction). 
Fig b) is a $1D$ plot of the field intensity along the line in this plane containing 
the direction of electric field polarization.}}
\label{fig_field_map}
\end{figure}

One of the principal sources of interest of the localized plasmon resonances is 
their capacity to produce large field enhancements in regions much smaller than 
the incident field wavelength. This property is demonstrated in 
fig.\ref{fig_field_map}a) with a 2D and 1D
plot of the electric field intensity in and near an isolated 
$50$ nm diameter silver sphere 
illuminated near its resonance peak ($\lambda_{0}=365\ \mathrm{nm}$ with 
$N_{\mathrm{Ag}}=0.077+1.6i$). The plots in Fig.\ref{fig_field_map} are performed in 
a plane containing the center of the sphere and perpendicular to 
$\mathbf{k}_{\mathrm{inc}}$ (the polarization lies along the horizontal axis). 
The dimensionless extinction and scattering `efficiencies', 
$Q=\sigma/(\pi R^{2})$, at this frequency are  are respectively $Q_{\mathrm{ext}}=14.48$ and 
$Q_{\mathrm{scat}}=6.76$.

\begin{figure}[ht]
\centering\includegraphics[width=0.7\textwidth]{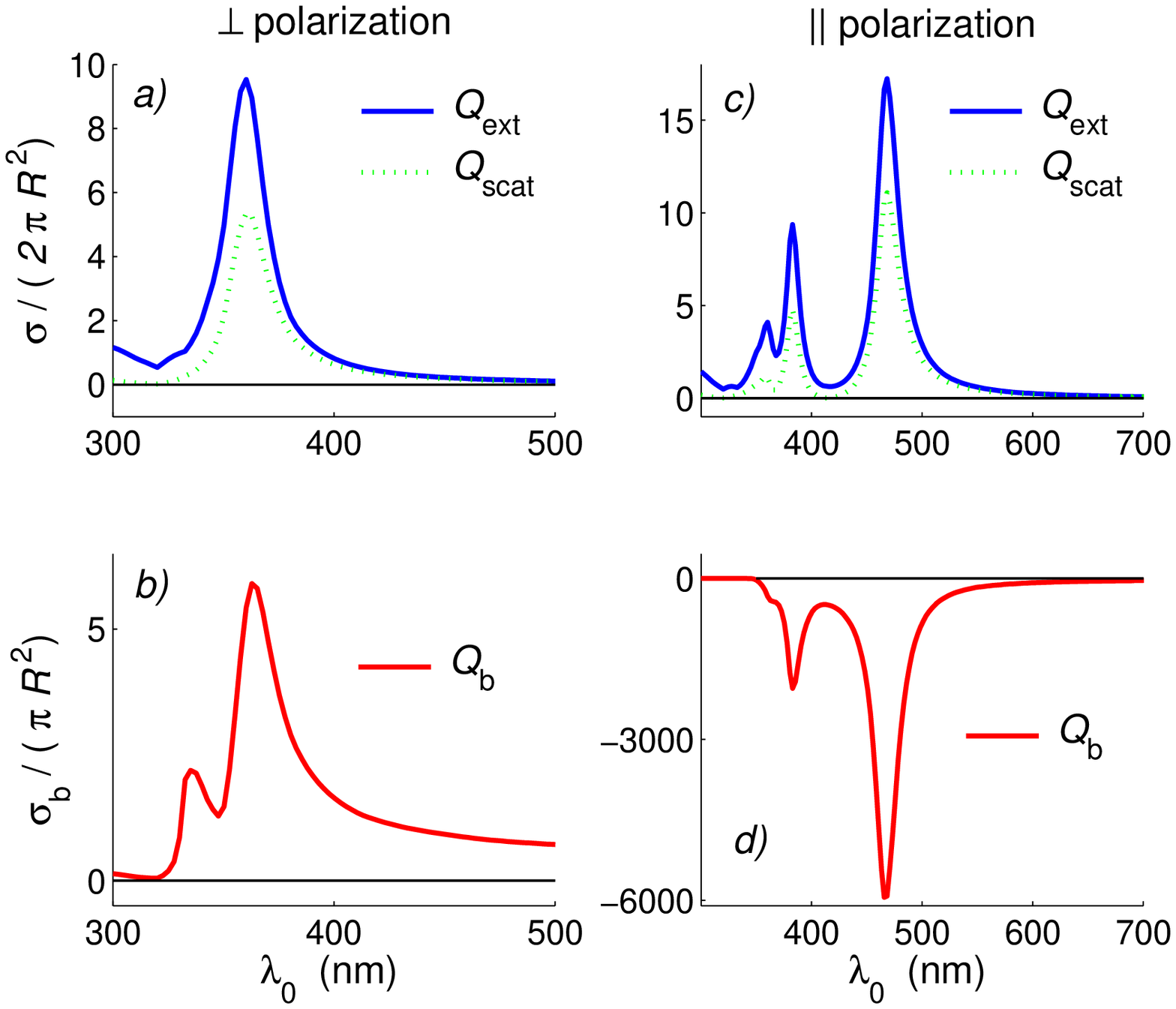}
\caption{\footnotesize Dimensionless cross section `efficiencies' per particle, $Q=\sigma/(2\pi R^{2})$ and binding
force `efficiencies' for a dimer of 50$\ \mathrm{nm}$ diameter spheres (1 nm separation). In
a) and b) the polarization is perpendicular to the symmetry axis and in c) and
d) it is parallel to the symmetry axis.}
\label{fig_2_part}
\end{figure}

We now use the balanced recursive technique to calculate the optical response
of a dimer composed of $50\ \mathrm{nm}$ diameter silver spheres whose surfaces are separated
by $1\ \mathrm{nm}$. Although the $T$ matrix calculated by RCTMA contains
information for arbitrary incident fields, we study the physically interesting case 
of a plane wave perpendicular to the axis separating the particles. As is widely known, 
the response then depends strongly on the polarization of the incident light. In 
figures \ref{fig_2_part}a) and \ref{fig_2_part}c), the extinction and scattering
cross sections per particle are presented when the polarization is respectively 
perpendicular and parallel to the symmetry axis. From figure \ref{fig_2_part}c), 
one sees that the cross section for the parallel to axis polarization presents a 
two sphere coupled resonance that is strongly red-shifted with 
respect to the isolated particle resonance. The polarization perpendicular to this axis
on the other hand presents only slight modifications with respect to an isolated sphere.

The optical binding force cross sections for these same
polarizations are respectively plotted in figures \ref{fig_2_part}b) and \ref{fig_2_part}d).
While the binding force for the polarization perpendicular 
to the particle axis (cf.\ref{fig_2_part}c)) is slightly repulsive, the force for polarization parallel
to the resonance can be highly attractive with the dimensionless
$\left\vert Q_{\mathrm{b}}\right\vert $ attaining amplitudes of three orders of magnitude.
There has already been experimental and theoretical evidence supporting the existence of
optical force couplings in particles with plasmon excitations\cite{li08} although such
high precision calculations at such small separations seems not to have
been presented before now.

This dimer system dramatically illustrates the `strong' coupling category 
since correct calculations require that the VSWF space be enlarged far beyond the 
predominantly dipolar response characterizing the particles in isolation. 
The normalized cross sections per particle are given in the table \ref{tab_1} for different
values of the VSWF space truncation. Although it was necessary to go to $\sim n_{\max}=30$ 
to achieve $4$ digit precision in all the cross sections, the table indicates that results were already 
quite good at $n_{\max}=20$.

{\footnotesize
\begin{table}[ht]
\centering
\begin{tabular}
[c]{l|cccccccc}
$n_{\max}$ & \qquad 5 \qquad & \qquad 10 \qquad & \qquad 15 \qquad & \qquad 20 \qquad & \qquad 25 \qquad & \qquad 30 & \qquad 35 \qquad &
\qquad 40 \qquad \\\hline
$Q_{\mathrm{ext}}/2$ & \ \ \ \ 4.60 & \ \ \ \ 15.53 & \ \ \ \  17.38 & \ \ \ \  17.20 & \ \ \ \  17.14 & \ \ \ \ 17.13 & \ \ \ \  17.13 &
\ \ \ \ 17.13\\
$Q_{\mathrm{scat}}/2$ & \ \ \ \   3.51 & \ \ \ \ 10.62 & \ \ \ \ 11.30 & \ \ \ \  11.04 & \ \ \ \  10.98 & \ \ \ \  10.97 & \ \ \ \ 10.97 &
\ \ \ \ 10.97\\
\ \ $Q_{\mathrm{b}}$ & \ \ \ \  -417 & \ \ \ \ -3639 & \ \ \ \ -5530 & \ \ \ \ -5918 & \ \ \ \ -6000 & \ \ \ \ -6015 & \ \ \ \ -6018 &
\ \ \ \ -6018
\end{tabular}
\caption{\footnotesize Dimensionless cross section `efficiencies' per particle in function
of the VSWF truncation, $n_{\mathrm{max}}$. $Q_{\mathrm{ext}}=\sigma_{\mathrm{ext}}/(2\pi R^{2})$,
$Q_{\mathrm{scat}}=\sigma_{\mathrm{scat}}/(2\pi R^{2})$ and $Q_{\mathrm{b}}=
\sigma_{\mathrm{b}}/(\pi R^{2})$. The system is a dimer composed of 
$D=50\ \mathrm{nm}$ diameter silver spheres (1 nm separation) at 
($\lambda_{0}=467\ \mathrm{nm}$ and $N_{\mathrm{Ag}}=0.048+2.827i$)}
\label{tab_1}
\end{table}}

\begin{figure}[ht]
\centering\includegraphics[width=1.0\textwidth]{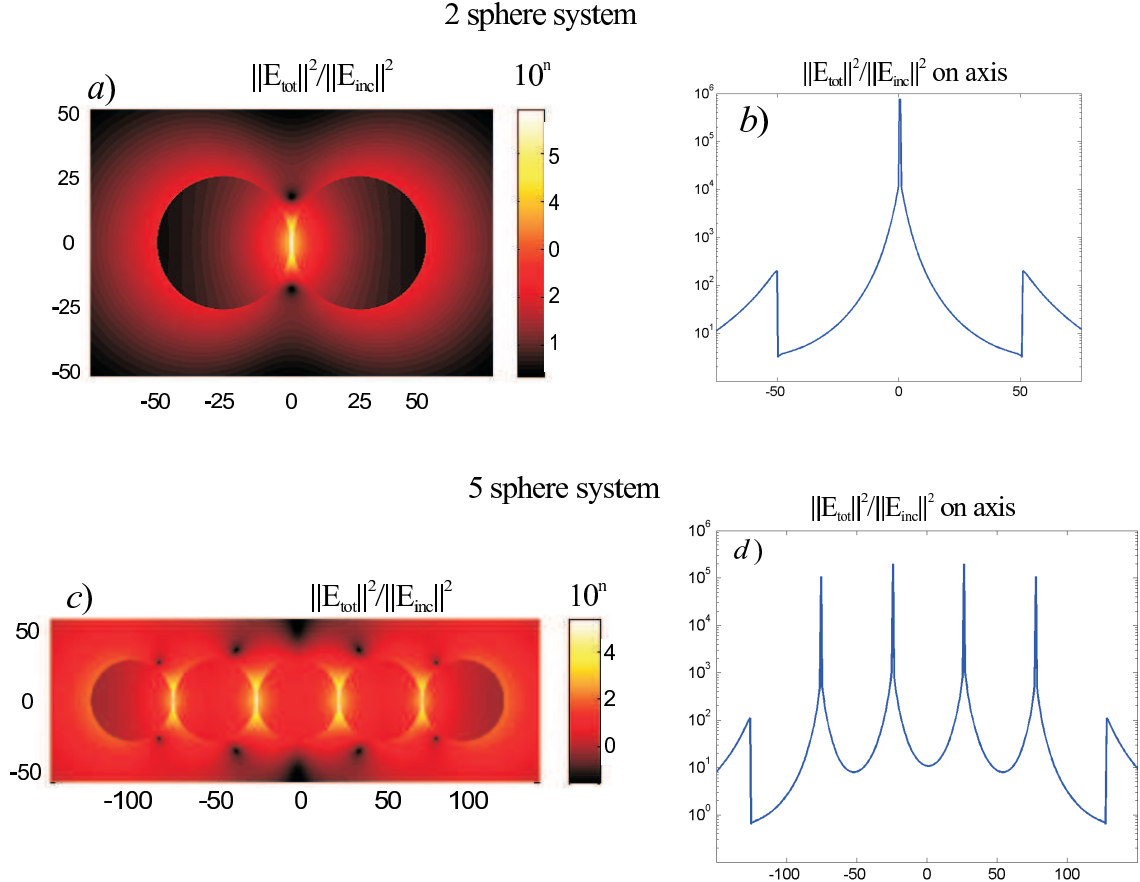}
\caption{\footnotesize Logarithmic scale plots of the field intensity for a two sphere dimer 
with ($\lambda_{0}=467\ \mathrm{nm}$ and incident light polarized along the sphere 
axis ($N_{\mathrm{Ag}}=0.048+2.827i$). 
In a) is a 2D plot in the plane containing the centers of the spheres and the polarization vector 
while b) is a 1D Logarithmic plot along the symmetry axis of the spheres. Figures. c) and d) are the 
same as a) and b) respectively but for a 5 sphere chain of spheres at its resonance maximum
($\lambda_0=561\ \mathrm{nm}$ and $N_{\mathrm{Ag}}=0.0564+3.685i$). (cf. figure \ref{fig_5_part}). }
\label{fig_fieldmaps}
\end{figure}

A base 10 logarithmic intensity field map for a two silver sphere dimer illuminated with light
polarized along the symmetry axis (frequency near the coupled sphere resonance maximum 
($\lambda_{0}=467\ \mathrm{nm}$ and $N_{\mathrm{Ag}}=0.048+2.827i$) is presented in 
figs.\ref{fig_fieldmaps}a) and figs.\ref{fig_fieldmaps}b) which are respectively a 2D plot 
(in the same plane as figure \ref{fig_field_map}) and a 1D plot along the symmetry axis. 
The size parameter of the individual spheres is $kR=0.34$ and the
isolated cross sections at this frequency are $Q_{\mathrm{ext}}=0.136$ and $Q_{\mathrm{scat}}=0.0963$.
As can be seen in fig.\ref{fig_fieldmaps}, the fact that one had to go so far beyond the
dipolar response has a dramatic effect on the field inside and near the
the particles. Notably, the fields inside the particles are no longer
quasi-constant as was the case for isolated particles.

An important word of caution should be made at this point. Although
$1\ \mathrm{nm}$ separation may appear to `nearly' touching, the coupled
resonance is in fact quite sensitive to exact separation details when resonant
particles are so closely separated. For example, at a separation distance of
$0.5\ \mathrm{nm}$ for the silver dimer, the coupled plasmon resonance is
displaced to $\lambda_{0}\simeq516\ \mathrm{nm}$ as compared with 
$\lambda_{0}\simeq467\ \mathrm{nm}$ for a $1\ \mathrm{nm}$ separation, and the multipole
order has to be pushed to $n_{\max}\simeq50$ to achieve four digit accuracy in
the cross sections. Nanometer scale separations are not necessarily
theoretical idealizations however as recent experiments with DNA separators
have demonstrated\cite{bi08}. Nevertheless, in applications like DNA separators, one
may well have to consider the strong optical forces between these
particles on account of the exceptionally strong attractive optical forces
efficiencies of these resonances. For instance, the binding force efficiency
at $0.5\\mathrm{nm}$ separation was calculated at $Q_{\mathrm{b}}=-20174$ for
$\lambda_{0}\simeq515.6\ \mathrm{nm}$ (cf. $Q_{\mathrm{b}}=-6018$ for $1\ \mathrm{nm}$
separation at $\lambda_{0}\simeq467\ \mathrm{nm).}$ The question of perfect
spheres exactly in contact however seems untenable from an experimental
standpoint and quite difficult from a theoretical standpoint on account of the
singular behavior of the contact point. Theoretical `separations' of
$1\mathring{A}$ for instance require multipole truncations of the order of
$n_{\max}\gtrsim120$ before convergence is achieved, but the idea of `perfect'
spheres separated by atomic scales has clearly gone beyond domain of
applicability of our mesoscopique physical model in any case.

\begin{figure}[ht]
\centering\includegraphics[width=0.9\textwidth]{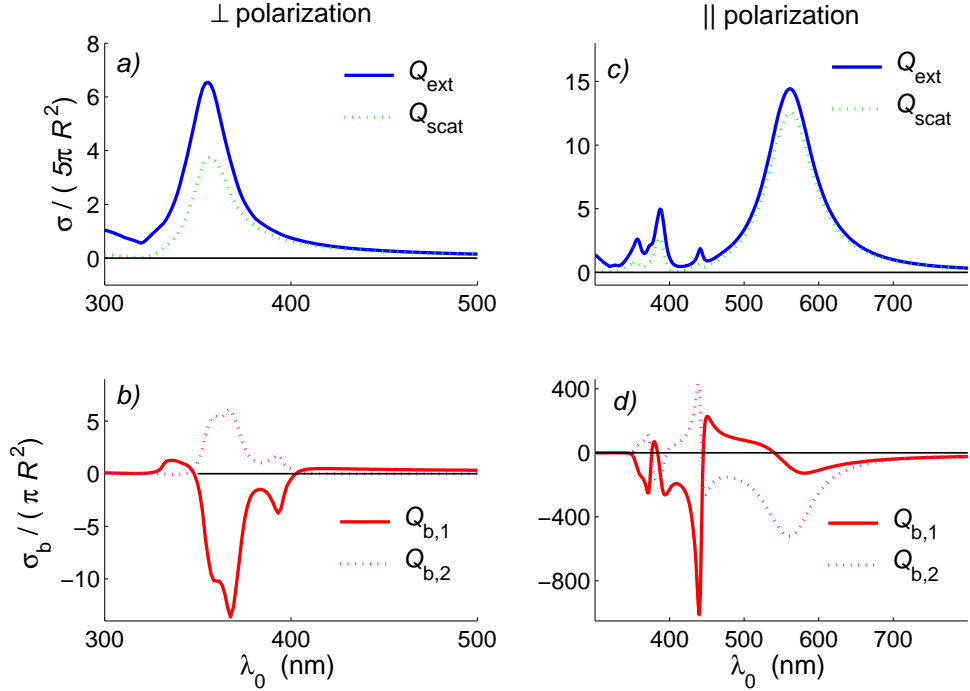}
\caption{\footnotesize Total cross section and binding `efficiencies' for a line of 5 `touching' silver
spheres 50 $\mathrm{nm}$ in diameter (1 nm separation). In a) and b) the polarization is
perpendicular to the symmetry axis and in c) and d) it is parallel to the
symmetry axis.}
\label{fig_5_part}
\end{figure}

It is also important to verify that the recursive algorithm works for more complicated
systems. Towards this end, we illustrate in figure \ref{fig_5_part}, the results of 
calculations for a system composed a line of $5$ identical silver spheres spheres 
separated by $1\ \mathrm{nm}$. For the binding force, we now present
$Q_{\mathrm{b},1}$ which is the binding optical force between outermost spheres and their
nearest neighbor and $Q_{\mathrm{b},2}$ which is the binding force between central sphere
and each of its nearest neighbors. It is interesting to remark that
addition of other spheres in the chain dramatically lessens the strong binding force interactions
between spheres even though the fields between the spheres (cf. fig.\ref{fig_2_part}) can still be 
almost as high as the dimer case.

We remark that the interactions have continued to red-shift and widen the coupled ``chain'' 
resonance. This chain resonance peaks at $\approx 561\ \mathrm{nm}$ and 
$N_{\mathrm{Ag}} \simeq 0.0564+3.685i$ ($Q_{ext}/5 = 14.416$ and $Q_{scat}/5=12.543$).
It is clear from figure \ref{fig_5_part}c) that the extinction cross section of
the chain resonance is increasingly dominated by scattering rather than absorption. 
The number of VSWF orders necessary for high precision was also seen to decrease slightly
for the chain. The calculation of fig.\ref{fig_5_part} was carried out with 
$n_{max}=20$ since calculations at $n_{max}=24$ produced negligible differences on this scale.

Despite the dominance of scattering, a considerable amount of absorption is
still present in the $5$ sphere chain. Furthermore, from the field maps in
figures \ref{fig_fieldmaps}c) and \ref{fig_fieldmaps}d), one can see that
the central sphere has the highest field internal field intensities, and one 
consequently expects increased absorption in the central sphere. This supposition
can readily be confirmed quantitatively by using eq.(\ref{abs}) to calculate
the absorption in each individual sphere. The results are given in table \ref{abstab}.

{\footnotesize
\begin{table}[ht]
\[
\begin{array}
[c]{|c|c|c|c|c|}\hline
Q_{\mathrm{a,1}} & Q_{\mathrm{a},2} & Q_{\mathrm{a},3} & Q_{\mathrm{a},4} &
Q_{\mathrm{a},5}\\\hline
0.8346 & 2.333 & 3.030 & 2.333 & 0.8346
\\ \hline
\end{array}
\]
\caption{\footnotesize Individual absorption efficiencies  $Q_{\mathrm{a},j}\equiv \sigma_{\mathrm{a},j}/(\pi R^2)$
in a five sphere chain at $\lambda_0=561\ \mathrm{nm}$ and $N_{\mathrm{Ag}}=0.0564+3.685i$}
\label{abstab}
\end{table}}

\begin{figure}[hb]
\centering\includegraphics[width=0.5\textwidth]{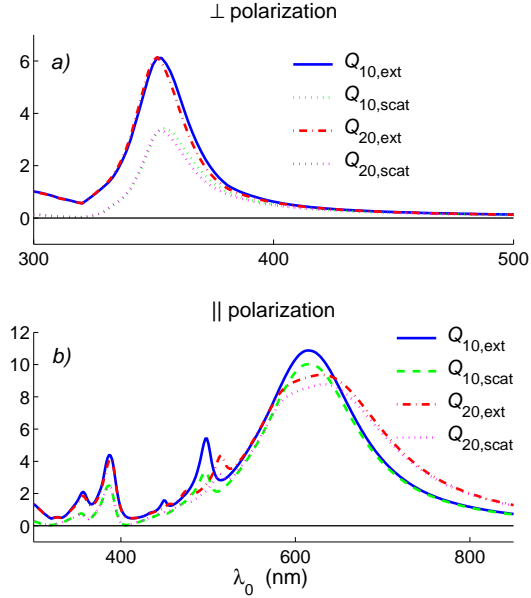}
\caption{\footnotesize Total extinction and scattering cross section efficiencies per particle in chains of 10 and 20 
particles. In a) the polarization is
perpendicular to the symmetry axis and in b) it is parallel to the
symmetry axis.}
\label{fig_20_part}
\end{figure}

We conclude this section with some calculations systems concerning larger chains 
of particles. One can remark that the chain coupled resonance continued to red-shift and 
widen when passing from the dimer to the five particle chain. Results for 
the extinction and scattering cross sections chains of 10 and 20 sphere 
chains are presented in figure \ref{fig_20_part} for the same polarizations and incident
directions as considered previously.
 
One readily sees that ultraviolet and perpendicular responses per particle seem to have stabilized
for large chains. The collective chain response on the other hand continued to broaden and slightly
red shift as one passed from $10$ to $20$ sphere chains and it is an interesting point for
future studies to examine the evolution of this phenomenon for even longer chains and to study
the impact of defaults in the chains.

\section{\label{Concl} Conclusions}

The balanced recursive algorithm can give useful and highly accurate
information in systems with large numbers of strongly interacting resonances.
This has been demonstrated herein for the case of localized plasmon resonances
and the studies presented here suggest that chains of closely spaced localized plasmons
can have potentially interesting applications with respect to  
frequency shifting and broadening. Although not demonstrated here, this technique also 
proves useful for treating closely spaced systems possessing surface resonances of 
`whispering gallery' type. 

It is worth remarking that matrix balancing seems to be a useful method to employ in almost any 
Foldy-Lax equation solution scheme, be that for direct system matrix inversion, iterative techniques
or linear system solutions. In fact, some modern matrix inversion programs actually integrate
numerical matrix balancing into their algorithms. Nevertheless, since the matrix balancing in Foldy-Lax equations
can be obtained analytically at relatively low computational cost, it seems beneficial to carry out this
balancing explicitly rather than relying on purely numerical treatments.

The matrix balanced RCTMA has potentially interesting applications for
other kinds of resonance phenomenon, notably whispering Gallery modes. Such studies are currently underway. 
Furthermore, the ability of the matrix balanced RCTMA to study defaults and small modifications in 
large complicated systems is particularly promising and will be employed in subsequent studies.

Brian Stout and Alexis Devilez would like to thank Ross McPhedran, Evgeny
Popov and Nicolas Bonod for helpful discussions. This work was funded in part by
the grant ANR-07-PNANO-006-03 "ANTARES" of the French National Research Agency.

\appendix

\section{\label{VSH}Vector spherical wave functions}

The vector spherical wave functions can be readily written in terms of the
Vector spherical harmonics (VSHs) and outgoing spherical \emph{Hankel} functions :
\begin{align}
\mathbf{\Psi }_{1,p}\left( k\mathbf{r}\right) & \equiv \mathbf{M}_{nm}(k\mathbf{r})
\equiv h_{n}^{+}\left( kr\right) \mathbf{X}_{nm}(\theta ,\phi ) \notag \\ 
\mathbf{\Psi }_{2,p}\left( k\mathbf{r}\right) & \equiv \mathbf{N}_{nm}
(k\mathbf{r})\equiv \frac{1}{kr}\left[ \sqrt{n\left( n+1\right) }
h_{n}^{+}\left( kr\right) \mathbf{Y}_{nm}(\theta ,\phi )+\left[
krh_{n}^{+}\left( kr\right) \right] ^{\prime }\mathbf{Z}_{nm}(\theta ,\phi )
\right]   \label{MetN}
\end{align}
In the same manner, the regular VSWFs are obtained by replacing the spherical 
Hankel functions in eq.(\ref{MetN}) by spherical \textit{Bessel} functions.
Our adopted definition of the VSHs is
\begin{equation}
\mathbf{Y}_{nm}(\theta,\phi)  \equiv\widehat{\mathbf{r}}\,Y_{nm} (\theta,\phi) \qquad
\mathbf{Z}_{nm}(\theta,\phi) \equiv\frac{r\bm{\nabla}Y_{nm}(\theta,\phi)}{\sqrt{n(n+1)}} 
\qquad \mathbf{X}_{nm}(\theta,\phi) \equiv\mathbf{Z}_{nm}(\theta,\phi)\wedge\widehat{\mathbf{r}}
\end{equation}
where the $Y_{nm}(\theta,\phi)$ are the scalar spherical harmonics. 



\end{document}